\documentstyle[sprocl]{article}

\def\Journal#1#2#3#4{{#1} {\bf #2}, #3 (#4)}

\def\NPB{{\em Nucl. Phys.} B}
\def\PLB{{\em Phys. Lett.}  B}

\def\PRD{{\em Phys. Rev.} D}

\def\be{\begin{equation}}
\def\ee{\end{equation}}
\def\bea{\begin{eqnarray}}
\def\eea{\end{eqnarray}}

\begin{document}

\begin{flushright}
UM-TH-98-06 \\
ITP-SB-98-44 \\
16 June 1998\\ 
\end{flushright}

\medskip

\title{A NOVEL FACTORIZATION FOR $F_L$ IN THE LARGE $x$ LIMIT}

\author{R. Akhoury, M. G. Sotiropoulos\footnote{Talk presented at the Sixth 
International Workshop on Deep Inelastic Scattering and QCD, Brussels, 1998.}}

\address{Randall Laboratory, University of Michigan, Ann Arbror MI 48109, USA} 

\author{G. Sterman}

\address{Unstitute for Theortical Physics, State University of New York \\
	Stony Brook, NY 11794, USA}


\maketitle\abstracts{ 
A novel factorization formula is presented for the longitudinal structure 
function $F_L$ near the elastic region $x \rightarrow 1$ 
of deeply inelastic scattering. 
In moment space this formula can resum all contributions to $F_L$ that are 
of order $\ln^k N/N$.   
This is achieved by defining a new jet function which probes the transverse 
momentum of the struck parton in the target at leading twist.  
The anomalous dimension $\gamma_{J^\prime}$ of this new jet operator 
generates in moment space the logarithmic enhancements coming from the 
fragmentation of the current jet in the final state.  
It is also shown how the suggested factorization for $F_L$ is related to 
the corresponding one for $F_2$ in the same kinematic region. }

\section{The $F_L$ factorization formula}
Experimental data on the ratio $R$ of the longitudinal over the transverse 
cross section in D.I.S. \cite{flexpt} suggest that the theoretical 
prediction to ${\cal O}(\alpha_s^2)$ \cite{DDKS,SG,ZvN} 
underestimates $F_L$ in the large $x$ region even with target 
mass corrections included. 
This may be indicative of large dynamical 
higher twist effects that can be parametrized in the context of O.P.E. 
In such a case it is useful to also study at the level of leading twist 
the resummed contributions from large logarithms $\ln(1-x)$ that do appear 
in the perturbative expansion of $F_L$ near the elastic region 
$x \rightarrow 1$. This is the question considered here. 

The analogous problem for $F_2$ is rather well understood 
\cite{Sterman,CatTre,KorMar}.  
The leading corrections there arise from the presence of terms 
like $\ln^k(1-x)/(1-x)_+$ in the coefficient functions of the O.P.E. 
The corresponding factorization formula in momentum  ($x$) space reads 
\begin{equation}
F_2(x, Q^2, \epsilon) = |H_2(Q^2)|^2 J \otimes V \otimes \phi,  
\label{f2fac}
\end{equation}
where $\otimes$ denotes the usual convolution in the longitudinal 
momentum fraction. 
$|H_2(Q^2)|^2$ is the short distance dominated hard scattering function. 
$V$ is the soft radiation function that contains all the enhancements coming 
from on-shell propagation of low frequency partons. This factor is universal 
among all D.I.S. observables and will enter as is in the corresponding $F_L$ 
formula.  $\phi$ is the parton distribution function that contains all the 
singularities (the only ones that are present) from initial state 
fragmentation. This factor is target specific. Non-singlet contributions 
are dominant in the $x \rightarrow 1$ region, so only scattering of quarks 
will be considered here. Finally, $J$ is the jet function for a stream of 
nearly collinear partons with total invariant mass ${\cal O}((1-x)Q^2)$. 
The definition of $J$ in terms of UV renormalized effective operators is 
\begin{equation}
J= F.T. \langle T \, \Phi_v(0, -\infty) \,\psi(0) \bar{\psi}(y)\, 
\Phi_{-v}(-\infty, y) \rangle/ V, 
\label{jdef}
\end{equation}
with $F.T.$ denoting Fourier transformation in momentum space and $\Phi_v$ 
the Wilson line operator along the $v$ light cone direction. 
All soft enhancements are removed from the jet by the denominator $V$. 

The main difference between $F_2$ and $F_L$ can be readily traced in their 
corresponding definitions as projections of the hadronic tensor 
$W_{\mu \nu}$. Specifically, for a massless quark target
\begin{equation}
F_L(x, Q^2,\epsilon) = \frac{8 x^2}{Q^2} p^\mu p^\nu 
W_{\mu \nu}(p, q,\epsilon).  
\label{fltens} 
\end{equation}
This means that $F_L$ will start at ${\cal O}(\alpha_s)$, and to this order it
is regular and dependent on the transverse momentum of the incoming quark. 
Since the factor $V \otimes \phi$ is common to all structure functions 
as $x \rightarrow 1$, a new jet function is needed in the factorization 
formula that takes into account the above special feature of $F_L$. 
It can be shown \cite{ASS} that this new jet function is 
\begin{equation}
J^\prime =  \left(\frac{1}{4\pi} \, \frac{8 x^2}{Q^2} \right) \, 
F.T. \langle \Phi_{v}(0, -\infty) \, / \! \! \! \!  D_\perp \psi(0) \cdot 
/ \! \! \! \! D_\perp\bar{\psi}(y) \Phi_{-v}(-\infty, y) \rangle/ V,
\label{Jprimedef}
\end{equation}
and that the corresponding factorization formula for $F_L$ in $x$ space is 
\begin{equation}
F_L(x, Q^2, \epsilon) = |H_L(Q^2)|^2 J^\prime \otimes V \otimes \phi.
\label{flfac}
\end{equation}
The main argument for the above factorization comes from the fact that 
all logarithmic enhancements in $F_L$ originate from the same characteristic 
regions in momentum space (pinch surfaces) as for any other observable 
in massless perturbation theory. The factorization theorem classifies 
these logarithmic enhancements according to origin, 
(fragmentation, soft or initial state) and generates each class from the 
UV renormalization of an effective non-local operator, like $J^\prime$ above.

\section{Sudakov resummation}
Once factorization is established in momentum ($x$) space, exponentiation 
in moment ($N$) space follows from the renormalization of the effective 
operators in the Mellin transformed factors $\tilde{J^\prime}$, $\tilde{V}$ 
and $\tilde{\phi}$. As expected in this kinematic regime, double logarithms 
from small angle soft emission are captured by the Sudakov or cusp anomalous 
dimension \cite{KodTrent} 
$\gamma_K = C_F \alpha_s/\pi + {\cal O}(\alpha_s^2)$. 
Collinear logarithms from 
the fragmentation of the current jet are captured by the anomalous dimension 
$\gamma_{J^\prime}$, which is novel and characteristic of $F_L$. 
To first order it is computed to be \cite{ASS}
\begin{equation}
\gamma_{J^\prime}(\alpha_s) = \frac{\alpha_s}{\pi} 
\left[ \frac{9}{2}C_F -2 C_A - 4 \zeta(2) \left(C_F-\frac{C_A}{2}\right)
\right] +{\cal O}(\alpha_s^2). 
\label{gammaJtotno}
\end{equation}
The Sudakov exponentiated form of $F_L$ can be written as 
\begin{eqnarray}
&\ &  \tilde{F}_L(N, Q^2, \epsilon) \, = \,
\frac{1}{N} \, \tilde{J}^\prime(\alpha_s(Q^2/N)) \, 
(\tilde{V} \cdot \tilde{\phi})(1/N, \alpha_s(Q^2), \epsilon) 
\nonumber \\
&\ &\times 
\exp \left[-\frac{1}{2} \int_{Q^2/N}^{Q^2} 
\frac{d \mu^2}{\mu^2} \left( 
\ln \frac{Q^2}{\mu^2} \gamma_K(\alpha_s(\mu^2)) 
+ \gamma_{J^\prime}(\alpha_s(\mu^2)) \right)	
 \right] 
\nonumber \\
&\ &
+ {\cal O}\left(\frac{\ln^0N}{N} \right), 
\label{flsol}
\end{eqnarray}
with boundary condition 
$\tilde{J}^\prime(\alpha_s) = C_F \alpha_s/\pi + {\cal O}(\alpha_s^2)$.

It is worth empasizing that terms which are power suppressed as $1/N$ in 
moment space can be resummed via the above procedure. 
Such power suppresed terms are leading in $F_L$.  
Note also that the jet function $J^\prime$ starts at ${\cal O}(\alpha_s)$.
The overall $\alpha_s$ factor introduces an ambiguity in the normalization 
of the perturbative expansion for $F_L$.
This feature is more reminiscent of partonic elastic scattering 
rather than electroweak scattering. 
Finally, it can be shown that Eqs.~(\ref{gammaJtotno}, \ref{flsol}) agree 
with the fixed order calculation of the O.P.E. coefficient function to 
${\cal O}(\alpha_s^2)$.

\end{document}